\documentclass[preprint,12pt]{elsarticle}
\usepackage[utf8]{inputenc}
\usepackage[T1]{fontenc}

\usepackage{amssymb}

\journal{Journal of Alloys and Compounds}

\begin{document}

\begin{frontmatter}



\title{Silver-Platinum nanoparticles and nanodroplets supported on silica surfaces: structure and chemical ordering}


\author{F. Ait Hellal}  
\author{C. Andreazza-Vignolle}   
\author{P. Andreazza}    
\author{J. Puibasset}   

\affiliation{organization={ICMN, CNRS, Universit\'e d'Orl\'eans},
            addressline={1b rue de la F\'erollerie, CS 40059}, 
            city={Orl\'eans},
            postcode={45071 cedex 02}, 
            country={France}}

\begin{abstract}
Stable and metastable metallic nanoparticles exhibit unique properties compared to the bulk, with potentially important applications for catalysis. This is in particular the case for the AgPt alloy that can exhibit the ordered L1$_1$ structure (alternation of pure Ag and Pt (111) planes) in nanometer size particles. However, for such small systems, the interfaces play an important role. Therefore, the support used to elaborate the nanoparticles in ultrahigh vacuum experiments may influence their properties, even in the case of weakly interacting substrates like amorphous carbon or silica. This work focuses on the AgPt nanoparticles deposited on silica, and investigates the effect of the support disorder and roughness on the structure and chemical ordering, in particular at the interface with the substrate, by Monte Carlo calculations of the atomic density profiles with semi-empiric potentials. 

\end{abstract}




\end{frontmatter}


\section{Introduction}

Supported metallic nanoparticles (NPs) are catalyst models for structure and chemical ordering studies \cite{RN1670, RN1666, RN1358, RN1663, RN2978}. Choosing an amorphous substrate like amorphous carbon or silicium oxide, as silica, has the advantage of minimizing the interactions with the NP, hence preserving the structure and morphology it would adopt in vacuum in the absence of interactions. This strategy is used in experiments where the NPs are grown on amorphous supports in ultrahigh vacuum conditions \cite{RN1456, RN1364, RN1374}. It has however been observed that this type of support may still influence the NP structure and morphology \cite{RN1778, RN1374}. Although the expected effects are less spectacular than for crystalline supports like MgO which strongly interact with NPs \cite{RN1432, RN1597, RN1370, RN1601, RN1599, RN1368, RN1568}, theoretical works have recently focused on the effect of weak surfaces \cite{RN2951, RN1764, RN1940}.

Among the possible effects of the substrate, chemical ordering is particularly relevant for catalysis, since NPs offers the unique opportunity to possibly drastically minimize the required amount of active matter by an optimal organization of the chemical species at the NP external surface.

The silver-platinum alloy exhibits interesting features \cite{RN1823, RN2981, RN1825, RN2833}, in particular an ordered L1$_1$ structure (alternation of pure Ag and Pt (111) planes) that has been observed in nanometer size particles \cite{RN2973, RN2345}, and stimulated theoretical studies \cite{RN2952, RN2950, RN2717}. It has also important applications as catalyst in fuel cells, where chemical ordering strongly influences its catalytic efficiency \cite{RN2976, RN2975, RN2974, RN2982}. Therefore, assessing the structure of the supported AgPt nanoalloy is a relevant issue. 

The elaboration process strongly influences the structure of the NP which is not necessarily at equilibrium (the system remains trapped in local minima, corresponding to metastable states) \cite{RN2345, RN3011}.
In simulations, the situation is even worse since the limited capabilities of computers impede to reach the experimental timescale, and thus strongly limits the possibilities of atomic reorganization.
To circumvent the problem, it is possible to increase the metal mobility by increasing the temperature \cite{RN2950}. This is why, besides the solid NP, we will also consider the corresponding liquid droplet at 1200 K and 1500 K.
Beyond the fact that this forces atomic mobility towards equilibrium, the disappearance of the L1$_1$ chemical order in the core due to the thermal agitation should leave room for the observation of a possible remnant chemical ordering at the interfaces, in particular that with the silica support.
It is emphasized that, in contrast to simulations, increasing the temperature of Ag-based nanoparticles like AgPt up to the liquid phase is experimentally difficult both in ultrahigh vacuum conditions, due to sublimation before melting, and at ambient pressure, due to the contamination by the atmosphere.

This paper is divided as follows: We first present the numerical model and the methods to characterize the structure and chemical ordering in terms of atomic density profiles. Then the results follow, with a focus on the effect of the temperature on the structure and chemical ordering, as well as the influence of the disorder and roughness of the support.

\section{Numerical Details and Methods}
{\bf Molecular model:} We perform Monte Carlo (MC) simulation of supported AgPt nanoparticles on various silica substrates that mimic the supports used to elaborate the systems in ultrahigh vacuum experiments. The system is described at the atomic level, with semi-empirical interatomic potentials. The simulations are performed in the canonical ensemble, including random displacements and atomic exchanges between the metallic species Ag and Pt. These exchanges are particularly useful at low temperature where the energy barrier associated to atomic diffusion in the core of the NP is too high to allow chemical reorganization.

\begin{figure}[]
\centering
\includegraphics[width=0.8\textwidth]{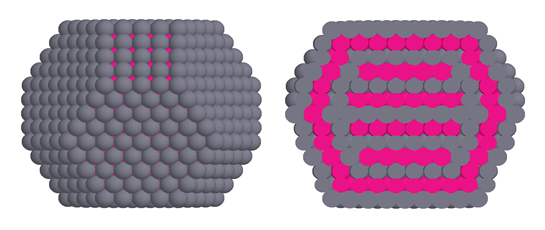}
\caption{\label{fig_NPs} Atomic configuration of the most stable TOh AgPt NP comprising 1289 atoms. Grey: silver; Magenta: platinum. Left: front view. Right: sectional view.}
\end{figure}

{\bf Interatomic potentials:} The many-body metal-metal interaction derives from the tight binding scheme in the second moment approximation (TBSMA) \cite{RN1442}. The ordered AgPt L1$_1$ structure being stabilized by the contribution of the second neighbors, an additional Gaussian term has been developed by Front and Mottet to reproduce the main structural properties of AgPt alloys \cite{RN2952, RN2950}. They have in particular determined the most stable structures of TOh AgPt NPs smaller than 7 nm. In this study we will focus on the NP with 1289 atoms (3.4 nm) which exhibits a rich structure depicted in Fig.~\ref{fig_NPs}. 

The metal-support interaction being weak (van der Waals like), a simple Lennard-Jones potential has been used, that has been previously developed for pure silver and platinum NPs as well as AgPt nanoalloys \cite{RN2977}. This potential has been parameterized to reproduce the aspect ratio (defined as H/D where H is the height and D the diameter) of experimentally deposited NPs (the parameters are given in Table~\ref{table1}). 

\begin{table}
\caption{Metal-silica support interaction, described by the Lennard Jones potential $V(r)=4\epsilon_{\rm M-S} \left[ \left( \sigma_{\rm M-S}/r \right)^{12} - \left( \sigma_{\rm M-S}/r \right)^6 \right] $ where M is the metal (Ag or Pt) and S is one of the silica species (Si or O) \cite{RN2977}.}
\begin{tabular}{c c c c c}
\hline\hline
 metal & $\epsilon_{\rm M-Si}$ (meV) & $\sigma_{\rm M-Si}$ (nm) & $\epsilon_{\rm M-O}$ (meV) & $\sigma_{\rm M-O}$ (nm) \\ [0.5ex]
\hline
 Ag   & 13  & 0.329  & 80  & 0.278  \\
 Pt   & 18  & 0.324  & 110 & 0.273  \\ [1ex]
\hline
\end{tabular}
\label{table1}
\end{table}

\begin{figure}[]
\centering
\includegraphics[width=0.8\textwidth]{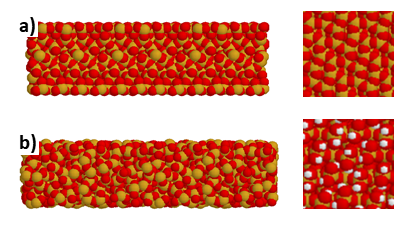}
\caption{\label{fig_supports} Side (left) and top (right) views of the two silica substrates. a) The (100) quartz surface (with the $1\times 2$ reconstruction visible on the top view). b) The amorphous SiO$_2$ substrate (the surface hydroxyls are visible on the top view). The lateral dimension is 6.8 nm.}
\end{figure}

{\bf Supports:} We consider two silica supports to evidence the effect of the atomic disorder and roughness that can be observed in experimental supports like oxidized silicon wafers. As a perfectly flat and ordered surface we use the (100) quartz surface. It is emphasized that this surface undergoes a $1\times 2$ reconstruction with a top layer twice as dense as the bulk quartz (see Fig.~\ref{fig_supports}a) \cite{RN1718}. To model a disordered substrate we simply cut and relax a slab of amorphous silica (a-SiO$_2$) (see Fig.~\ref{fig_supports}b). More details in the method and the potentials used can be found in Ngandjong et al. \cite{RN1764, RN1940}. It is mentioned that, despite the fact that the amorphous silica surface is hydroxylated, the hydrogen species are not explicitly taken into account in the interactions. 

{\bf Chemical ordering and density profiles:} The objective is to measure the effect of the substrate on the chemical ordering in the nanoalloy. This is done simply by comparing layer by layer with the equilibrium structure of the free NP (Fig.~\ref{fig_NPs}) since the geometric structure is marginally affected by the weak interaction with the substrate. However, the time scale involved in experiments being inaccessible to simulations, the intrinsic mobility of Ag (due to its lower cohesion) is largely underestimated in the calculations. The introduction of MC exchanges between Ag and Pt greatly solves the problem and allows chemical rearrangement, but possibly misses some facets of the complex cross-diffusion of Ag and Pt in the NP. We therefore consider the effect of temperature, ranging from the solid up to the liquid state (above approximately 1200K for Ag$_3$Pt) \cite{RN2950}. In this case, the NP structure is fully disordered (droplet) but may exhibit partial chemical ordering, in particular close to the interfaces (with vacuum and substrate). 

\begin{figure}[]
\centering
\includegraphics[width=0.9\textwidth]{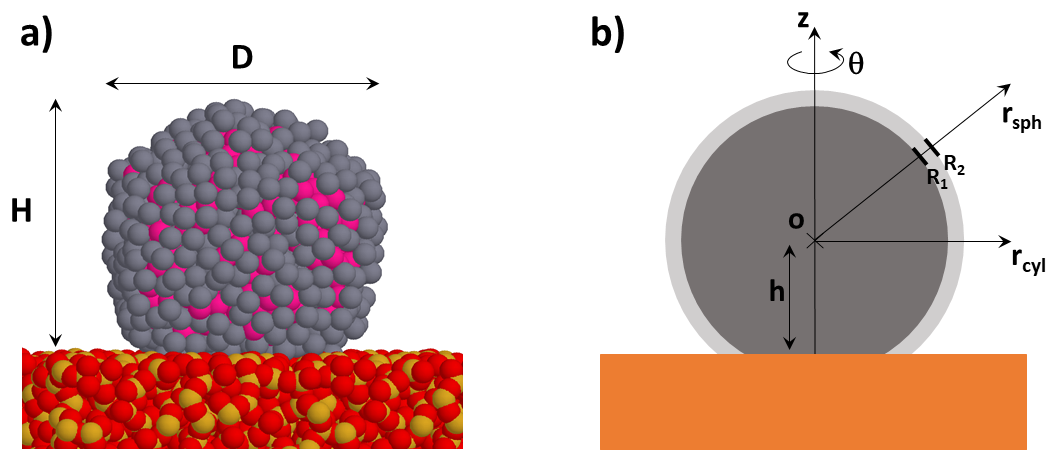}
\caption{\label{fig_drop} Snapshot of a AgPt droplet on the amorphous silica at $T$ = 1500 K. The aspect ratio is defined as $H/D$ where $H$ is the height of the droplet above the substrate and $D$ its diameter. b) Simple model of a supported and truncated liquid drop with uniform density in the core (<$R_1$) and a thin skin with smoothly varying density between $R_1$ and $R_2$. $r_{cyl}$, $\theta$ and $z$ are the cylindrical coordinates with the origin at the center of the droplet; $r_{sph}$  is the spherical radial distance from the origin.}
\end{figure}

In the liquid case, the structure of the nanodroplet is characterized by averaging the atomic density profiles of the metal. Since our objective is to determine the aspect ratio ($H/D$) we focus on two density profiles, along the $z$ axis (perpendicular to the surface) and along the radial coordinate $r_{cyl}$  (in the plane parallel to the substrate, see Fig.~\ref{fig_drop}). So, from the local atomic density $\rho(r_{cyl},\theta,z)$ we construct the two quantities:
\begin{eqnarray*}
\xi_r (z) = \int \rho(r_{cyl},\theta,z) r_{cyl} dr_{cyl} d\theta \\
\xi_z (r_{cyl}) = \int \rho(r_{cyl},\theta,z) r_{cyl} dz d\theta.
\end{eqnarray*}
These density profiles have the dimension of the inverse of a distance, and their integrals give the total number of atoms in the NP. 

In order to determine the aspect ratio from the data, these profiles are fitted on a simple model of a truncated sphere of uniform density $\rho_0$ of radius $R_1$ with a skin of thickness $R_2-R_1$ where the density decreases from $\rho_0$ down to zero. We define the droplet radius as $R=0.5(R_1+R_2)$ and the aspect ratio is $H/D=(h+R)/2R$ (see Fig.~\ref{fig_drop}). In practice, $\xi_z (r_{cyl})$ mostly depends on the droplet radius, while $\xi_r (z)$ is sensitive to $h$.

\section{Results}

\subsection{Structure and chemical ordering of solid AgPt NPs on quartz}

\begin{figure}[]
\centering
\includegraphics[width=0.6\textwidth]{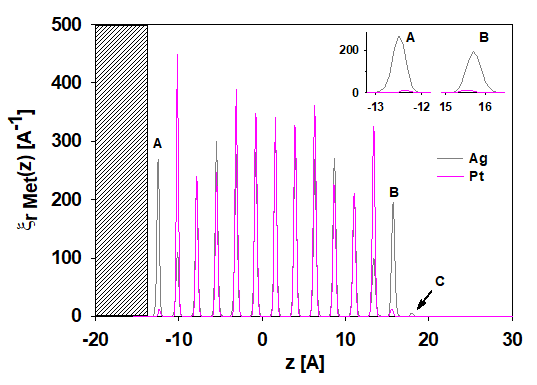}
\caption{\label{fig_prof_quartz_700} Density profiles $\xi_r (z)$ for each metal species in AgPt on quartz at $T$ = 700~K. Inset: zoom of the first and last peaks.}
\end{figure}

At low temperature (around room $T$ or below), the supported AgPt NP remains essentially frozen in its initial state, preserving its highly structured layers and chemical ordering (in particular the L1$_1$ structure) with only scarce exchanges between Ag and Pt species. Increasing the temperature enhances the atomic mobility and somehow mimics the experimental conditions where moderate annealing allows the AgPt NPs to reorganize thanks to the large mobility of Ag atoms. The optimal temperature is around 700~K, the largest possible below the melting point of Ag for a NP of few nanometers \cite{RN1401}. We first consider the AgPt NP deposited on the perfectly ordered quartz surface. The atomic density profiles along the $z$ axis are acquired for Ag and Pt and shown in Fig.~\ref{fig_prof_quartz_700}.

\begin{figure}[]
\centering
\includegraphics[width=0.5\textwidth]{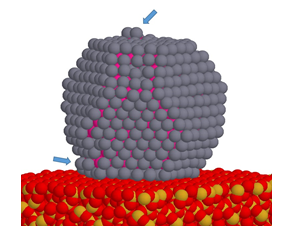}
\caption{\label{fig_snap_quartz_700} Snapshot of the AgPt NP on quartz at $T$ = 700 K showing adatoms on the top layer and close to the substrate (arrows).}
\end{figure}

One observes a strong layering through the whole NP showing that at 700~K the NP remains solid during the simulation run. One can however observe some atomic mobility at the external surface of the NP, as revealed by the small peak C in Fig. 4 corresponding to adatoms on the top layer. An example where such adatoms can be seen is given in Fig.~\ref{fig_snap_quartz_700}.

Despite the low atomic mobility, the MC chemical exchanges between Ag and Pt species allow the system to reorganize. It is observed that the chemical ordering associated to the alternating Ag and Pt planes in the core of the NP is lost, showing that the L1$_1$ structure is destabilized by the temperature. 

On the other hand, the outermost silver layer is quite robust, as can be seen on the snapshot (Fig.~\ref{fig_snap_quartz_700}) and the presence of essentially pure Ag peaks in the first and last layers denoted A and B in Fig. 4. Note however that these layers are not perfectly pure Ag anymore, revealing that some Pt atoms can diffuse in the outer Ag shell. But, as can be seen in the inset, the peaks associated to Pt in these layers are not centred with respect to the corresponding Ag peaks. In each case the Pt peak is shifted towards the centre of the NP, meaning a strong penalty for Pt at the surface. Quantitatively, the integration of Pt peaks gives their proportion in the A and B layers. One gets for A 3.7\% Pt and for B 5.5\% Pt. The free Ag surface is slightly more favourable for Pt compared to the surface in contact with the support. This is an interesting feature at odds with the observation that the Pt-SiO$_2$ interaction is stronger than the Ag-SiO$_2$ interaction (Table~\ref{table1}). A possible interpretation is that the Ag layer at the interface with silica is more constrained than the free one.

\subsection{Structure and chemical ordering of AgPt nanodroplets on quartz}

What happens above the melting point of AgPt? The structure of the NP is now expected to be destabilized (and at equilibrium thanks to the atomic mobility), which could influence the chemical ordering. The double objective is thus to determine the aspect ratio of the AgPt droplet and the chemical profiles at the interfaces. The first calculations are done well above the melting point, at $T$ = 1500~K, for the AgPt NP deposited on the quartz surface.

\begin{figure}[]
\centering
\includegraphics[width=\textwidth]{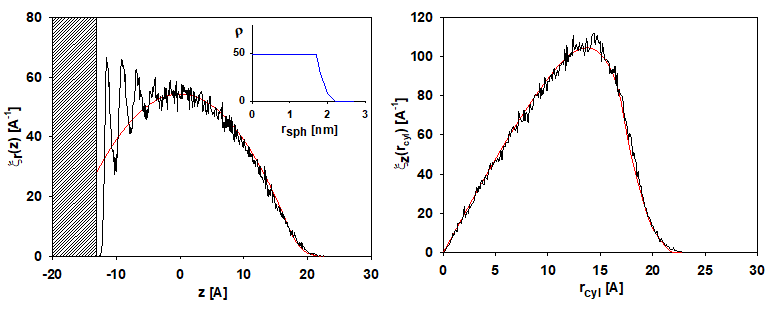}
\caption{\label{fig_prof_quartz_1500} Atomic density profiles $\xi_r (z)$ (left panel) and $\xi_z (r_{cyl})$ (right panel) for the AgPt droplet comprising 1289 atoms on the quartz support at $T$ = 1500~K. The Ag and Pt atoms are not distinguished in this figure. The grayed region in the left panel for $z$<-13~{\AA} represents the quartz support. The smooth red lines are best fit with the liquid drop model (Fig.~\ref{fig_drop}b). Inset: density profile used for the liquid drop model.}
\end{figure}

We first calculate the density profiles $\xi_r (z)$ and $\xi_z (r_{cyl})$ for all atoms in the drop without distinguishing Ag and Pt (Fig.~\ref{fig_prof_quartz_1500}). As can be seen on $\xi_r (z)$, the layer structure along the $z$ axis is smoothed out due to the thermal agitation, except close to the perfectly flat and ordered quartz surface, where one can observe at least three layers. In the upper region of the droplet the layering has completely disappeared and the density profile decreases smoothly with a small tail at $z$ = 20~\AA .

Along the radial coordinate, $\xi_z (r_{cyl})$ shows an initial increase essentially linear corresponding to the integration of a uniform density on the surface of a cylinder. It then reaches a maximum and rapidly decreases due to the spherical shape of the drop, with a tail due to the smooth transition between the metal core and the surrounding vacuum. 

Examination of the density profiles gives a good estimate of the drop height and diameter, but a best fit with the liquid drop model depicted in Fig.~\ref{fig_drop}b gives a better insight (smooth solid red lines in Fig.~\ref{fig_prof_quartz_1500}). Obviously, the layering observed in $\xi_r (z)$ cannot be described by the uniform density model, but the average variations are caught. However, this reveals that the first layers are particularly structured because of the height of the maxima compared to the drop model curve. This is essentially because of the perfectly ordered quartz surface. Otherwise, the model describes quantitatively the variations of $\xi_r (z)$ far from the support, as well as the variations of $\xi_z (r_{cyl})$ in the whole range of values. The corresponding density profile of the liquid drop model is shown in the inset, and the values given by the best fit are $h$ = 13~{\AA} and $R$ = 19.5~{\AA}, giving the aspect ratio $H/D$ = 0.83. The uniform density is $\rho_0$ = 0.049 atom/{\AA}$^3$.
 
\begin{figure}[]
\centering
\includegraphics[width=0.6\textwidth]{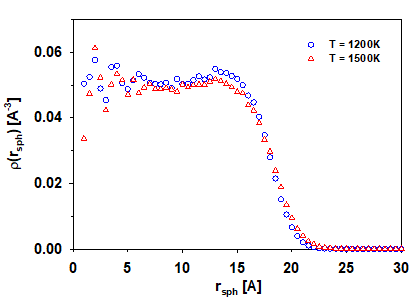}
\caption{\label{fig_radial_quartz} Radial (spherical coordinate $r_{sph}$) atomic density profiles for the AgPt droplet comprising 1289 atoms on the quartz support at $T$ = 1200 K and 1500~K. The Ag and Pt atoms are not distinguished in this figure.}
\end{figure}

The density profile in the inset shows that the thickness of the skin (5~{\AA}) somehow corresponds to 1 to 2 atomic diameters and is not negligible compared to the radius of the droplet. A more refined profile can be acquired directly during the course of the simulation by calculating a spherical radial distribution $\rho(r_{sph})$ taking care to exclude the rays in the solid angle defined by the intersection between the sphere and the substrate. The result is given in Fig.~\ref{fig_radial_quartz}. It confirms that the density within the core of the droplet is essentially uniform within uncertainties, due to the low statistics in the centre of the sphere. The average density on the plateau is in agreement with the value previously extracted from the best fit. It also confirms that the atomic density profile drops smoothly to zero within a skin thickness of 5~{\AA}. 

\begin{figure}[]
\centering
\includegraphics[width=\textwidth]{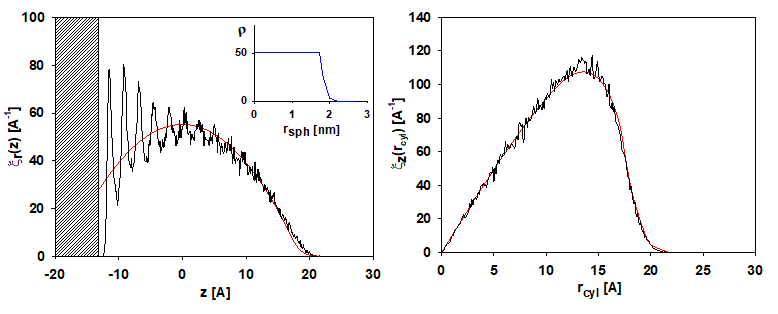}
\caption{\label{fig_prof_quartz_1200} Same as Fig.~\ref{fig_prof_quartz_1500} for $T$ = 1200~K.}
\end{figure}

The same analysis has been done at a lower temperature $T$ = 1200 K (see Fig.~\ref{fig_prof_quartz_1200} for $\xi_r (z)$ and $\xi_z (r_{cyl})$ and Fig.~\ref{fig_radial_quartz} for $\rho(r_{sph})$). As can be seen, reducing the temperature enhances the observed layering at the quartz surface. Otherwise, the structure is not affected significantly except for a slightly larger density in the core: $\rho_0$ = 0.051 atom/{\AA}$^3$. The extracted values from the fit are $h$ = 13~{\AA} and $R$ = 19.7~{\AA}, giving an aspect ratio $H/D$ = 0.83.

\begin{figure}[]
\centering
\includegraphics[width=0.6\textwidth]{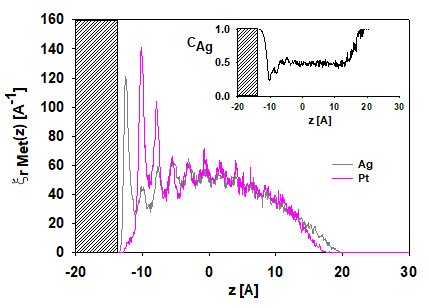}
\caption{\label{fig_prof_AgPt_quartz_1200} Density profiles along the $z$ axis $\xi_r (z)$ for each metal species in AgPt on quartz at $T$ = 1200~K. Inset: Ag concentration.}
\end{figure}

In order to quantify the chemical ordering at the interface due to the support we focus only on the case of the AgPt droplet at 1200~K on quartz: the low temperature and the strong interaction with the support is expected to enhance the effect. Figure~\ref{fig_prof_AgPt_quartz_1200} shows that at midheight (around $z$=0) the Ag and Pt species have equal probabilities to be at any position. However, close to the substrate, one observes a strong chemical ordering with an almost pure Ag layer at the interface with the quartz, the second layer being filled essentially with Pt, the silver atoms being at the periphery. On the other side (top of the NP), we also observe chemical ordering (silver excess at the interface with vacuum). All this is explained by the low surface tension of Ag which preferentially migrates to the interfaces, while Pt accumulates in subsurface, in particular at the interface with the support, a behaviour similar to what was observed in the solid state.

\subsection{Effect of the support disorder and roughness on the liquid NPs}

\begin{figure}[]
\centering
\includegraphics[width=\textwidth]{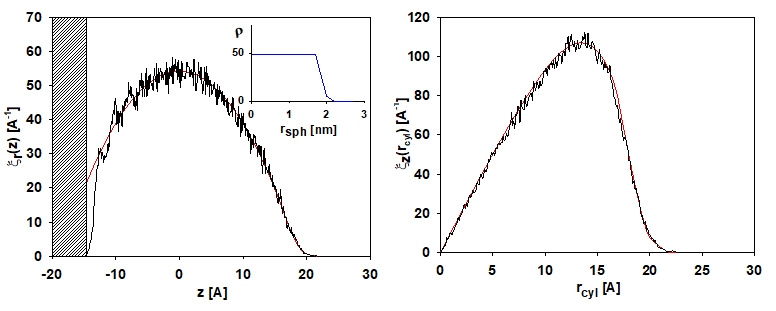}
\caption{\label{fig_prof_silica_1500} Same as Fig.~\ref{fig_prof_quartz_1500} where the quartz support is replaced by the amorphous silica (see Fig.~\ref{fig_supports}b). Inset: density profile used for the liquid drop model. The temperature is $T$ = 1500 K.}
\end{figure}

Does the strong layering observed on the quartz persist in presence of disorder or roughness? To answer this question, we performed MC simulations of the AgPt NP on the amorphous SiO$_2$ surface (Fig.~\ref{fig_supports}b) at $T$ = 1500 K. The density profiles $\xi_r (z)$ and $\xi_z (r_{cyl})$ exhibit essentially the same characteristics as for the quartz support, except for two points (see Fig.~\ref{fig_prof_silica_1500})

(i) The layering close to the support is significantly smoothed out due to the atomic disorder of the amorphous silica surface. Note that the roughness of this surface is quite small, but also participates to the destabilization of the first layer. The consequence is that the density profile now closely approaches the smooth curve given by the liquid drop model.

(ii) The best fit with the model gives the following parameters: $h$ = 14.5~{\AA} and $R$ = 19.5~{\AA}, giving an aspect ratio $H/D$ = 0.87. As can be seen, compared to quartz, the aspect ratio is slightly closer to 1, in agreement with the fact that the surface density of the a-SiO$_2$ is lower than that of the quartz which exhibits a densification due to the reconstruction. Otherwise, the density in the core of the drop is $\rho_0$ = 0.049 atom/{\AA}$^3$, a value identical to that observed on the quartz.

\begin{figure}[]
\centering
\includegraphics[width=\textwidth]{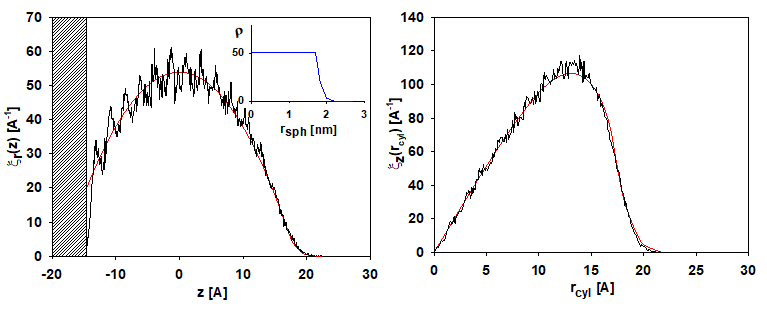}
\caption{\label{fig_prof_silica_1200} Same as Fig.~\ref{fig_prof_silica_1500} at $T$ = 1200~K.}
\end{figure}

Reducing the temperature favors layering along the $z$ axis (see Fig.~\ref{fig_prof_silica_1200}). It is however much less pronounced than on quartz. The main difference is that here the layering has a uniform amplitude from the first layer in contact with the substrate up to the center of the NP, while on quartz it was highly increasing in the vicinity of the surface. This suggests that in the case of the quartz support, the surface clearly imposes a strong ordering, while, on the a-SiO$_2$ support, the layering is essentially intrinsic to the metal structure although it is of course initiated and stabilized by the surface. 

The best fit with the drop model gives the following parameters: $h$ = 14.5~{\AA} and $R$ = 19.5~{\AA}, giving an aspect ratio $H/D$ = 0.87, and $\rho_0$ = 0.051 atom/{\AA}$^3$, a value identical to that on the quartz at the same temperature.

\section{Conclusion}

The structure and chemical ordering in AgPt NPs of 3.4~nm (1289 atoms) deposited on silica supports have been investigated by Monte Carlo simulations. The introduction of chemical exchanges between the Ag and Pt species allows to converge towards the chemical ordering at equilibrium even for NPs trapped in a metastable structure in terms of atomic positions. It is observed that silver preferentially migrates to the outer surface and at the interface with the substrate, preserving the same stable structure as for the free NP. Increasing the temperature to 700~K allows partial atomic mobility without melting the NP. One observes Ag adatoms on the external surface and close to the substrate, and a small diffusion of Pt atoms from the subsurface layer to the surface layer. 
It is however observed that the Pt atoms at the periphery always remain slightly embedded in the outermost Ag layer, which is of importance for catalysis. Since the local structure can be strongly influenced by the surrounding atmosphere, further studies are necessary to quantify the catalytic efficiency of this system in real conditions. Systems exhibiting similar structures or chemical ordering illustrate this point \cite{RN3007, RN3006, RN3004, RN3005}. 

Higher temperatures have been considered, in the liquid state. The layer structure is expected to disappear, but the presence of the support is able to stabilize the first layers. This is particularly visible for the ordered quartz surface, but less pronounced for the disordered amorphous silica. The determination of the aspect ratio shows that the morphology of the supported drop follows the expected behaviour, with a lower aspect ratio on the more attractive quartz surface. Regarding chemical ordering, the external surface at the interface with the substrate is mostly composed of Ag species, with statistically few Pt atoms. Here again, examination of density profiles shows that the Pt atoms always remain slightly embedded in the outermost Ag layer. 

\section{Declaration of interests}
The authors declare that they have no known competing financial interests or personal relationships that could have appeared to influence the work reported in this paper.

\section{Acknowledgements}
F. A. H. acknowledges a grant from the education and research ministry for her Ph.D. The authors would like to acknowledge support from the International Research Network - IRN ``Nanoalloys'' of CNRS.



\bibliographystyle{elsarticle-num} 
\bibliography{manuscript_AgPt}





\end{document}